\documentstyle[floats,epsfig,aps,prb]{revtex}
\newcommand{\be}{\mbox{e}}
\newcommand{\ve}{\mbox{\bf e}}
\newcommand{\dfrac}[2]{\frac{\strut \displaystyle{#1}}{\displaystyle{#2}}}

\begin{document}

%\draft

%%%%%%%%%%%%%%%%
\twocolumn[\hsize\textwidth\columnwidth\hsize\csname
@twocolumnfalse\endcsname

\title{\large\bf Magnetization plateaus of the Shastry-Sutherland
  model for SrCu$_2$(BO$_3$)$_2$:
  Spin-density wave, supersolid and bound states}

\author{Tsutomu Momoi\cite{address}}
\address{Lyman Laboratory of Physics, Harvard University, Cambridge,
  MA 02138}
\author{Keisuke Totsuka}
\address{Department of Physics, Kyushu University, Hakozaki,
  Higashi-ku, Fukuoka-shi, 812-8581 Japan}
\date{\hspace*{5cm}}
%\date{\today}

\maketitle

\begin{abstract}
We study the Heisenberg
antiferromagnet on the Shastry-Sutherland lattice under magnetic fields,
to clarify the magnetic properties of SrCu$_2$(BO$_3$)$_2$.
Treating magnetic excitations promoted by the field as Bose
particles and using strong coupling expansion, we derive an
effective Hamiltonian for the effective magnetic particles.
Anisotropic repulsive interactions between effective particles
induce `insulating' states with a stripe SDW structure at magnetization
$m/m_{\rm sat}=1/3$ and a checkerboard structure at 1/2, and thereby
form magnetization plateaus. 
Supersolid phases appear around insulating SDW phases by changing
the magnetic field. Nature of
these supersolid phases is discussed in detail. 
We also demonstrate how the geometry of the Shastry-Sutherland
lattice affects dynamical properties of magnetic excitations 
significantly and makes a novel type of
quintuplet ($S=2$) boundstates condense for very small magnetization.
\end{abstract}
\vskip-2mm
\pacs{PACS numbers: 75.60.Ej, 75.10.Jm}
%%%%%%%%%%%%%%%
]
\narrowtext

\section{Introduction}
Since plateau structures were observed in the
magnetization process of a series of
quasi one-dimensional Ni-compounds\cite{NarumiHSKNT},
magnetization plateaus have been attracting
extensive interests. The appearance of plateaus in magnetization
curves was explained as metal-insulator transitions of magnetic
excitations driven by a magnetic field\cite{Totsuka};
Magnetic excitations crystallize
and form SDW in the plateau states, and they are itinerant in the
non-plateau states. Recently it was discussed that this phenomenon is
not limited to the
one-dimensional systems, but more general, and occurs in two- and
three-dimensional systems as well\cite{Kolezhuk,MomoiSK,MomoiT}.
Before the recent studies, magnetization plateaus were
already known to appear
at $m/m_{\rm sat}=1/3$ in the antiferromagnetic compounds
on the triangular lattice, C$_6$Eu\cite{SuematsuOSSMD},
CsCuCl$_3$\cite{NojiriTM}, and RbFe(MoO$_{4}$)$_{2}$\cite{InamiAG}.
Theoretically, plateaus were seen at $m/m_{\rm sat}=1/3$ in the
Heisenberg antiferromagnet on the triangular
lattice\cite{NishimoriM,ChubukovG,JacobsNS}, and also at
$m/m_{\rm sat}=1/2$ in the multiple-spin exchange model with four-spin
interactions\cite{KuboM,MomoiSK}.
The 1/3-plateau comes from appearance of a collinear
{\it uud} state\cite{NishimoriM,ChubukovG,JacobsNS} and
the 1/2-plateau from the {\it uuud} state\cite{KuboM,MomoiSK}.
These magnetization plateaus can also be regarded as
superfluid-insulator transitions of flipped-spin degree of
freedom\cite{MomoiSK,MomoiT}.
It may be worth mentioning that there is also another trial to realize
magnetization plateau in
two-dimensional systems as gapped spin liquid states
analogous to the FQHE wave functions\cite{Liquid}.

\begin{figure}[tbp]
  \begin{center}
    \leavevmode
    \epsfig{file=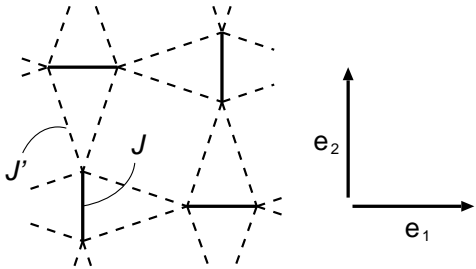,width=2.5in}
  \end{center}
  \caption{Shastry-Sutherland lattice. Solid (dotted) lines denote
    bonds with strong exchange $J$ (weak one $J'$).}
  \label{fig:lattice}
\end{figure}

Recently a quasi two-dimensional compound
SrCu$_2$(BO$_3$)$_2$ is
attracting extensive interests because it shows
magnetization plateaus and peculiar dynamical properties.
 The two-dimensional lattice structure of Cu$^{2+}$ ions in
SrCu$_2$(BO$_3$)$_2$ is the so-called Shastry-Sutherland
lattice\cite{SmithK,Kageyama}, which is shown in Fig.\
\ref{fig:lattice}. Susceptibility and specific heat data show that
interactions are antiferromagnetic, the spin
excitation has a gap above the ground state, and the spin anisotropy
is weak\cite{Kageyama}. This material seems\cite{MiyaharaUa}
to be well described with the $S=1/2$ Heisenberg
antiferromagnet on the Shastry-Sutherland lattice\cite{ShastryS}.
(Hereafter we call
this model simply as Shastry-Sutherland model.)
 The ground state of the Shastry-Sutherland model is exactly a
direct product of local dimer singlets on bonds $J$ for the region
$J'/J<0.68$ (Refs.~\onlinecite{ShastryS,MiyaharaUa}) and there is a
finite gap above the ground state. Susceptibility and specific heat
estimated from this model with $J'/J=0.68$ fit well with
the experimental results\cite{MiyaharaUa}.
(Recently the value has been modified to $J'/J=0.635$ by taking into
account the three-dimensional coupling of Shastry-Sutherland
layers.\cite{MiyaharaUb})
In ref.\ \onlinecite{Kageyama},
Kageyama et al.\ reported two plateaus
at $m/m_{\rm sat}=1/8$ and $1/4$ in the magnetization
curve of SrCu$_2$(BO$_3$)$_2$.
Theoretically, we studied the magnetization process of
the Shastry-Sutherland model, treating a dimer triplet as a particle,
and thereby predicted a novel
broad plateau at $m/m_{\rm sat}=1/3$ in the previous report\cite{MomoiT}.
It was argued that
the appearance is due to superfluid-insulator transition of the
excitations.
Quite recently, the above 1/3-plateau was experimentally observed in
magnetization measurements up to a strong field of 57[T]\cite{Onizuka}.
As was predicted in ref.\ \onlinecite{MomoiT}, this
plateau is the broadest one ever found in this material.
This seems to support the correctness of
our argument based on the particle picture.
This material also shows peculiar dynamical 
properties\cite{KageyamaB,Nojiri,Room,Raman}, e.g.
one-magnon excitation is almost dispersionless, but two-magnon excitations
have strong dispersion\cite{KageyamaB}.
The aim of this paper is to present the details of our analyses and
results reported briefly in ref.\ \onlinecite{MomoiT}, and to proceed
further thereby giving remarkable consequences of the
correlating hopping of the effective Hamiltonian. This correlated
hopping can also explain the peculiar dynamical behaviors
observed in experiments\cite{TotsukaMU}.

The ground state of the Shastry-Sutherland model at zero
magnetic field was studied for a varying ratio $J'/J$
by the mean field approximation\cite{AlbrechtM},
exact diagonalization method\cite{MiyaharaUa}, and series
expansion\cite{KogaKawakami}.
It was found\cite{KogaKawakami} that
the ground state is the exact dimer state for $J'/J<0.69$, a gapped
plaquette singlet state for
$0.69<J'/J<0.86$, and N\'eel-ordered state for $0.86<J'/J$. The parameters
of SrCu$_2$(BO$_3$)$_2$ estimated\cite{MiyaharaUa,MiyaharaUb} as
$J'/J=0.68$ (or 0.635) suggest that the spin state of the real material
belongs to the dimer phase, but it is very close to the phase boundary
with the plaquette singlet phase. Consistency between theoretical and
experimental results on magnetization plateau at
$m/m_{\rm sat}=1/3$\cite{MomoiT,Onizuka}
and dynamical behavior in inelastic neutron
scattering\cite{KageyamaB,TotsukaMU} also
supports that the real material is in the dimer phase.

In this paper, we study the $S=1/2$ Heisenberg antiferromagnet on the
Shastry-Sutherland lattice and discuss the
magnetic properties under a magnetic field.
We analyze this model using strong-coupling expansion.
In section \ref{sec:effec_H}, we derive an effective Hamiltonian for
the dimer-triplet excitations.
Virtual triplet excitations yield various effective
repulsive interactions, which are responsible for the plateaus.
Although the usual single-particle hopping is completely
missing from the resulting Hamiltonian,  correlated hopping processes
are contained instead.

In section \ref{sec:mag_pro},
we investigate the magnetization process 
applying the classical approximation to the effective
(pseudo-spin) Hamiltonian
and show that plateaus appear at $m/m_{\rm sat}=1/3$ and $1/2$.
Both of the plateau states are Mott (SDW) insulators of magnetic
excitations.   Spatially anisotropic interactions perturbatively generated
stabilize several SDW structures. Near the plateau
states, there are supersolid phases, in which SDW long-range order (LRO) and
superfluid coexist\cite{AndreevL,Leggett-70,LiuF,FisherN,Matsuda-T-80}. 
Field ($B$) versus coupling ($J^{\prime}/J$) phase diagram
is also presented.

Although the inter-dimer coupling $J^{\prime}/J$ is not small,
the special geometry of the Shastry-Sutherland lattice
strongly suppresses bare one-particle hopping
and hence the correlated hopping becomes important.
This fact leads to remarkable
consequences on the motion of magnetic excitations.
In section IV, we investigate correlated hopping more closely
and demonstrate how it favors the
formation of a bound pair of dimer triplets which {\em repel} each
other at the level of the bare Hamiltonian.
This bound state has a relatively large dispersion and
may be an elementary particle at very low magnetization,
instead of the single dimer triplet.
The non-plateau state would be superfluid
of these bound states at least for very small magnetization.
 Above a certain threshold value of magnetization,
individual dimer triplets become elementary particles and
the non-plateau state is characterized by superfluidity of single
dimer triplets.

In section \ref{sec:sat}, we also discuss spin excitation just below
the saturation field. It is found that lowest energy states exist on a 
close curve instead of a point in the momentum space.

Critical phenomena of the plateau transition are discussed
in section \ref{sec:critical} and are argued to be in the same
universality class as the
superfluid transition of the interacting boson system in the dilute
limit.

%%%%%%%%%%%%%%%%%%%%%%%%%%%%%%%%%%%%%%%%%%%
\section{Effective Hamiltonian}
%%%%%%%%%%%%%%%%%%%%%%%%%%%%%%%%%%%%%%%%%%%
\label{sec:effec_H}
In this section we derive an effective Hamiltonian for the magnetic
excitations under a strong enough magnetic field.
We begin with the $J'=0$ limit.
In this limit, the lowest triplet excitation over the (dimer)
singlet ground state is apparently obtained by promoting
one of the dimer singlets to a triplet.
Although the dimer product remains to be the exact ground state even
for non-zero $J'$, the above completely localized triplet does not; 
perturbation $J^{\prime}$ `broadens' the triplet by exciting
nearby singlets.   Unlike the ground state which is perfectly
free from quantum fluctuation, excited states strongly
suffer from it.   This is one of the most important features
of the model.
As a result, (physical) triplets can interact with each other
with the help of {\em virtual} triplets created by perturbation
and the effective Hamiltonian for the physical triplet
degrees of freedom should contain {\em effective} interactions.

The most systematic way to take into account such virtual processes
would be the strong-coupling expansion\cite{Totsuka,Kolezhuk}.
We start from the limit $J^{\prime}=0$ and
treat the interaction with coupling $J'$ by perturbation.
In the absence of the external field, the spin states of a single
isolated dimer ($J$) bond consists of a singlet and triplet separated
by a gap $J$.
When the field is increased until the lowest triplet with $S^z=1$
intersects the singlet,  we may keep only two states--%
the singlet and the lowest triplet--as the physical
degrees of freedom as far as the low-energy sector is considered.

We carry out the degenerate perturbation for such low-energy
degrees of freedom.
Considering the triplet ($S=1$) state with $S^z=1$
as a particle (a hard-core boson)
and the dimer singlet ($S=0$) as a vacancy, we derive
an effective Hamiltonian for the magnetic particles.
(Rest of spin states, i.e.\ triplets with $S^z=0$ and $-1$, are
included into the intermediate virtual states of the perturbation.)
The perturbational expansion is performed up to the 3rd order in $J'/J$
for degenerate states with a constant number of dimer triplet
excitations with $S^z=1$.
The final form of the effective Hamiltonian is as follows
\begin{eqnarray}
H &=& H_0 + H_1 + H_2 + H_3, \label{eqn:eff_Ham} \\
H_0 &=& (J-B)\sum_i n_i, \\
H_1 &=& \dfrac{J'}{2}
  \sum_{\langle i,j \rangle} n_i n_j, \\
H_2 &=& -\dfrac{J'^2}{J}\sum_i n_i
+ \dfrac{J'^2}{2J} \sum_{\langle i,j \rangle} n_i n_j \nonumber\\
&+& \dfrac{J'^2}{4J} \sum_{i\in A} \{
2 n_{i+\ve1}(1-n_{i})n_{i-\ve1}
+ (b_{i+\ve2}^\dagger b_{i-\ve2}+h.c.) n_{i}
\nonumber\\
& &
+ (b_i^\dagger b_{i+\ve2}- b_i^\dagger b_{i-\ve2} + h.c.)
(n_{i-\ve1}-n_{i+\ve1})
\} \nonumber\\
&+& \dfrac{J'^2}{4J} \sum_{i\in B} \{ \ \ \ \ve_1 \leftrightarrow \ve_2 \ \ \
\},
\\
H_3 &=&
% 1,8 & 6,13
-\dfrac{J'^3}{2J^2} \sum_{i} n_i
% 2,9,4,11 no ishibu
-\dfrac{J'^3}{8J^2} \sum_{\rm NN} n_{i} n_{j}
+\dfrac{J'^3}{4J^2} \sum_{\rm NNN} n_{i} n_{j}
\nonumber\\
% 6,13
&+& \dfrac{J'^3}{16J^2} \sum_{i \in A} [
  4 n_{i+\ve2} n_{i} n_{i-\ve2}
% 3
- 12 n_{i+\ve1} (n_{i} - 1) n_{i-\ve1}
\nonumber\\
% 6,13
& &+6 (b_{i+\ve2}^\dagger b_{i-\ve2} + h.c.) n_{i}
\nonumber\\
% 2,9,4,11 no ichibu
& &+ 5 (b_{i}^\dagger b_{i+\ve2} - b_{i}^\dagger b_{i-\ve2} + h.c.)
 (n_{i-\ve1} - n_{i+\ve1})
\nonumber\\
% 15,22,c,c'
& &- (b_{i+\ve2}^\dagger b_i + h.c.)
 (n_{i+\ve2+\ve1} n_{i-\ve1} - n_{i+\ve2-\ve1} n_{i+\ve1})
\nonumber\\
& &- (b_{i-\ve2}^\dagger b_i + h.c.)
 (n_{i-\ve2-\ve1} n_{i+\ve1} - n_{i-\ve2+\ve1} n_{i-\ve1})
\nonumber\\
& &-2 (b_{i+\ve2+\ve1}^\dagger b_i + h.c.) n_{i+\ve2} n_{i-\ve1}
\nonumber\\
& &-2 (b_{i-\ve2-\ve1}^\dagger b_i + h.c.) n_{i-\ve2} n_{i+\ve1}
\nonumber\\
& &-2 (b_{i-\ve2+\ve1}^\dagger b_i + h.c.) n_{i-\ve2} n_{i-\ve1}
\nonumber\\
& &-2 (b_{i+\ve2-\ve1}^\dagger b_i + h.c.) n_{i+\ve2} n_{i+\ve1}
\nonumber\\
% 17,21
& &+ (b_{i+\ve2}^\dagger b_i + h.c.)
 (n_{i+\ve2+\ve1} n_{i+\ve1} - n_{i+\ve2-\ve1} n_{i-\ve1})
\nonumber\\
& &+ (b_{i-\ve2}^\dagger b_i + h.c.)
 (n_{i-\ve2-\ve1} n_{i-\ve1} - n_{i-\ve2+\ve1} n_{i+\ve1})
\nonumber\\
& &+ 2 (b_{i+\ve2+\ve1}^\dagger b_i + h.c.) n_{i+\ve2} n_{i+\ve1}
\nonumber\\
& &+ 2 (b_{i-\ve2-\ve1}^\dagger b_i + h.c.) n_{i-\ve2} n_{i-\ve1}
\nonumber\\
& &+ 2 (b_{i-\ve2+\ve1}^\dagger b_i + h.c.) n_{i-\ve2} n_{i+\ve1}
\nonumber\\
& &+ 2 (b_{i+\ve2-\ve1}^\dagger b_i + h.c.) n_{i+\ve2} n_{i-\ve1}
\nonumber\\
% 16,A
& &+ 2 (b_{i+\ve2-\ve1}^\dagger b_{i-\ve2}
- b_{i+\ve2+\ve1}^\dagger b_{i-\ve2} + h.c.)
n_{i+\ve2} n_{i}
\nonumber\\
& &+ 2 (b_{i-\ve2-\ve1}^\dagger b_{i+\ve2}
- b_{i-\ve2+\ve1}^\dagger b_{i+\ve2} + h.c.)
n_{i-\ve2} n_{i}
\nonumber\\
& &- (b_{i+\ve2}^\dagger b_{i-\ve2} + h.c.) n_{i}
\nonumber\\
& & \mbox{\hspace{5mm}}\times (n_{i+\ve2+\ve1} + n_{i+\ve2-\ve1} 
       + n_{i-\ve2-\ve1} + n_{i-\ve2+\ve1})
\nonumber\\
% 18
& & - \{ (b_{i}^\dagger b_{i-\ve2} + h.c.) n_{i+\ve2}
       - (b_{i}^\dagger b_{i+\ve2} + h.c.) n_{i-\ve2} \}
\nonumber\\
& & \mbox{\hspace{1cm}}\times      (n_{i-\ve1} - n_{i+\ve1})
\nonumber\\
& &+ 2 (b_{i+\ve2}^\dagger b_{i-\ve2} + h.c.) n_{i} (n_{i-\ve1}
 + n_{i+\ve1})
\nonumber\\
% 20,B
& &- 3 (b_{i}^\dagger b_{i-\ve1} + h.c.) n_{i+\ve1} (n_{i-\ve2-\ve1}
- n_{i+\ve2-\ve1})
\nonumber\\
& &- 3(b_{i}^\dagger b_{i+\ve1} + h.c.) n_{i-\ve1} (n_{i+\ve2+\ve1}
- n_{i-\ve2+\ve1})]
\nonumber\\
%%%%%%
&+&
\dfrac{J'^3}{16J^2} \sum_{i\in B} [ \ \ \ \ve_1 \leftrightarrow \ve_2 \ \ \ ],
\end{eqnarray}
where indices $i$ and $j$ run over an effective square lattice 
of dimer bonds (both horizontal and vertical), and $A$ ($B$) sublattice
contains horizontal (vertical) ones. The operator $b^\dagger_i$
($b_i$) creates (annihilates) magnetic particle with spin $S^z=1$ at
bond $i$, and $n_i=b_i^\dagger b_i$.
The unit vectors $\ve_1$ and $\ve_2$ are shown
in Fig.~\ref{fig:lattice}. The interactions summed up on B
sublattices are obtained by replacing $\ve_1$ and $\ve_2$ in those on A
sublattices. The abbreviations NN and NNN denote pairs of nearest
neighbor and next nearest neighbor sites.

The Hamiltonian derived above does not have the bare one-particle
hopping terms like $b^{\dagger}_{i}b_{j}+b^{\dagger}_{j}b_{i}$,
which means a single dimer triplet excitation is
dispersionless at this order.
This is in a striking contrast with other spin gap systems
{\em e.g.} the 2-leg ladder.
The energy gap of one dimer triplet excitation
is evaluated as $\Delta E_1 (B)=J-B-J'^2/J-J'^3/(2J^2)$.

On the other hand, the effective Hamiltonian contains many
correlated-hopping processes, where an effective hopping of a particle
is mediated by another one.
Roughly speaking, these are closely related to 2-particle
Green functions of the triplet bosons.
This correlated hopping is important for
dynamics of the Shastry-Sutherland model and it leads to many
interesting conclusions; bound states without attraction,
supersolid with stripe structure, etc.
Most terms of higher orders concern the correlated hopping.

Many two-body repulsive interactions are also derived, whose range and 
geometry are shown in
Fig.~4 of Ref.~\onlinecite{MomoiT}. 
The nearest neighbor repulsion $V_{\rm NN}$ and 
the next-nearest-neighbor one $V_{\rm NNN}$ are derived as 
$V_{\rm NN}=J'/2+J'^2/2J-J'^3/8J^2$ and $V_{\rm NNN}=J'^3/4J^2$.
(There were a typographical 
error and a mistake in the coefficients of the 3rd order terms of
$V_{\rm NN}$ and $V_{\rm NNN}$ in ref.~\onlinecite{MomoiT}. 
There was also the same mistake in
ref. \onlinecite{MiyaharaU00}.) The 3rd-neighbor repulsion $V_{\rm 3rd}$
between vertical (horizontal) bonds with the distance $2\ve_1$
($2\ve_2$) is $V_{\rm 3rd}=J'^2/2J+3J'^3/4J^2$. Thus $V_{\rm 3rd}$ 
is anisotropic and 
acts only in one direction. 
The effective Hamiltonian (1) correctly reflects the 
space-group symmetry of the original Shastry-Sutherland model. 
If it is considered as a model of
interacting hard-core bosons on a {\em square} lattice consisting of
dimer bonds, it still contains matrix elements which are not invariant 
under naive $90^\circ$ rotation about one site. 
%Since the original lattice structure has low symmetry ($D^{11}_{2d}$),
%the resulting Hamiltonian does not have $90^\circ$ rotational invariance.
We show in section~\ref{sec:mag_pro} that they lead to anisotropic SDW 
states with a stripe structure.
Longer-range repulsions between particles can appear from
higher-order perturbations.
Actually, Miyahara and Ueda \cite{MiyaharaU00} independently
took into account such long-range interactions
in a phenomenological manner, also adding weak
{\em one-body} hopping by hand.
Our finding is that there is much stronger {\em correlated}
hopping processes and that triplets are not necessarily localized.

%%%%%%%%%%%%%%%%%%%%%%%%%%%%%%%%%
\section{Magnetization process}
%%%%%%%%%%%%%%%%%%%%%%%%%%%%%%%%%
\label{sec:mag_pro}
In order to discuss the ground state properties of the effective Hamiltonian
(\ref{eqn:eff_Ham}), we first consider only
repulsive interactions.
From naive consideration of the range and geometry of the repulsive
interactions (see Fig.~4 in Ref.~\onlinecite{MomoiT}),
we can imagine various insulating density-wave states of dimer
triplets. Let us treat the repulsive interactions obtained at different
orders of $J'/J$, separately.
\begin{itemize}
\item[(a)] The strongest interaction comes from the 1st-order term and it
is repulsive for adjacent triplets.
This repulsive interaction chooses SDW with
2-(dimer)sublattices checkerboard
structure at $m/m_{\rm sat}=1/2$, where the unit cell has 4 sites
in the original Shastry-Sutherland lattice.
(See Fig.\ 6(b) in Ref.\ \onlinecite{MomoiT}.)
\item[(b)] The next strong repulsive interactions originate from 2nd
order terms and they are contained in $V_{\rm NN}$ and $V_{\rm NNN}$. 
Because of anisotropic interaction $V_{\rm 3rd}$, a stripe SDW state with a
6-(dimer)sublattice structure is stabilized at $m/m_{\rm sat}=1/3$ 
as shown in Fig.\ \ref{fig:SDW0.3}.
\item[(c)] The 3rd-order perturbation generates the weakest 
interactions, which favor
SDW of 10-(dimer)sublattices checkerboard structure at $m/m_{\rm sat}=1/5$.
\end{itemize}
These SDW states do not necessarily appear if we consider all
interactions and correlated hopping terms together.
Whether SDW insulator is realized or not is
determined by competition between repulsive interactions
and correlated hopping.

To consider both effects of correlated hoppings and interactions
together, we study the effective Hamiltonian in the classical limit.
To this end, we map the hard-core boson system to
the $S=1/2$
quantum (pseudo)spin system and then approximate the Pauli matrices by
the components of a classical unit vector.
Within the mean-field approximation, we first search for the ground
state taking into account
two and four (dimer)sublattices with a checkerboard
structure and also three (dimer)sublattices with a stripe structure,
because insulating states with these configurations are expected
from the above consideration.
To take account of larger sublattice structures,
we next study the ground state of a finite system with a Monte Carlo
method, where we gradually decrease temperatures to zero.
We consider a finite cluster of 60 dimers and impose
a periodic boundary condition that matches with all SDW
configurations expected from the repulsive interactions,
i.e. 2-(dimer)sublattice and 5-(dimer)sublattice checkerboard
structures and 3-(dimer)sublattice stripe one.

\begin{figure}[tbp]
  \begin{center}
    \leavevmode
    \epsfig{file=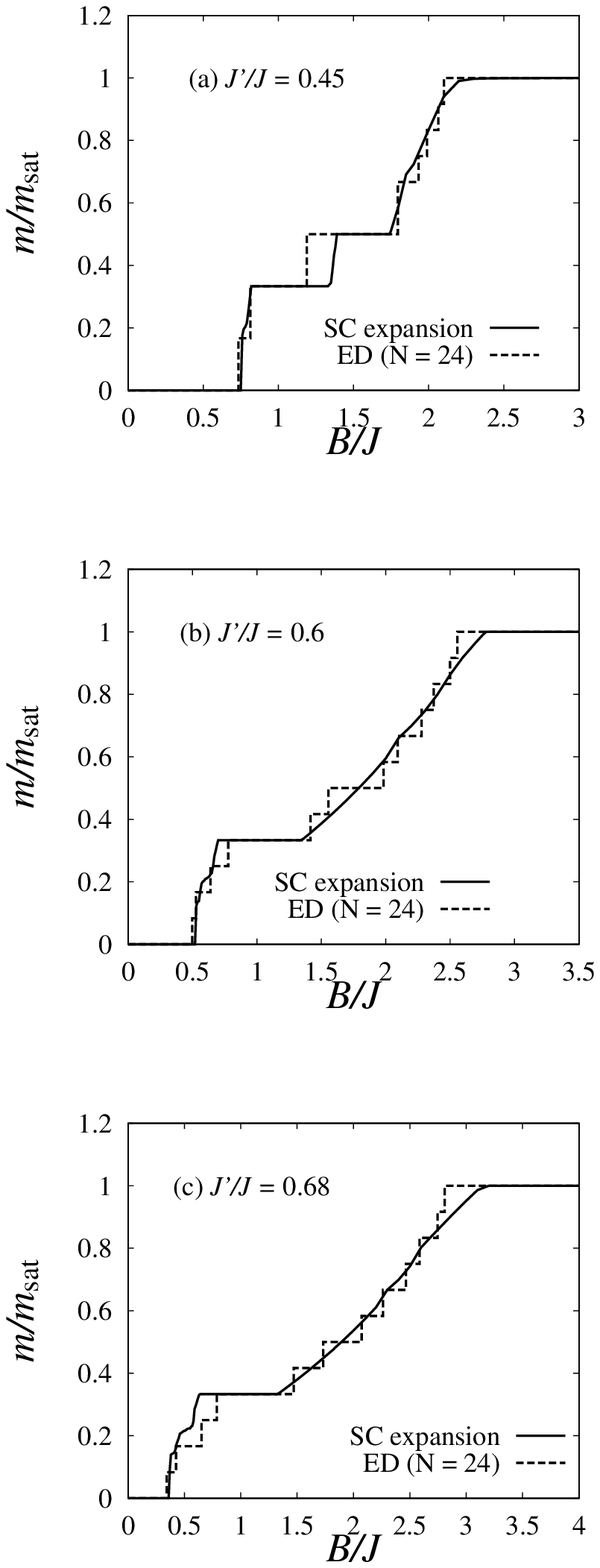,width=3.1in}
  \end{center}
  \caption{Magnetization processes of the Shastry-Sutherland model
   with $J'/J=0.45$ (a), $J'/J=0.6$ (b), and $J'/J=0.68$ (c).
   Solid line is obtained
  from the effective Hamiltonian in the classical limit and dotted one
  from the exact diagonalization (ED) of the original Shastry-Sutherland
  model with the size $N=24$. The solid line for $J'/J=0.68$ (c), 
  taken from Fig.~5 of Ref.~\protect\onlinecite{MomoiT}, is also shown 
  for the comparison with ED.}
  \label{fig:mag_pro}
\end{figure}

The evaluated magnetization processes are shown for the cases
$J/J'=0.45$, $0.6$, and $0.68$ in
Figs.~\ref{fig:mag_pro}(a), (b), and (c), respectively.
Plateau structures appear at magnetization $m/m_{\rm sat}=1/3$ and
1/2, but no plateau appears at $m/m_{\rm sat}=1/5$ and 1/4.
Near $m/m_{\rm sat}=1/4$, the slope of the curve becomes less steep,
but not flat. This means that the $m/m_{\rm sat}=1/4$ state is
energetically stable, though it does not have spin gap.
In the plateau phases at $m/m_{\rm sat}=1/3$ and 1/2, there are SDW
orders and no off-diagonal
long-range order (ODLRO), that is, collinear spin states are realized.
In the non-plateau phases, spins have
ODLRO. Spin configuration of each phase is explained in the following
subsections.
A remark is in order here about our classical approximation.
We obtained several plateaus by analyzing the classical pseudo-spin
Hamiltonian.   From this, one may conclude that these plateaus
are of a classical origin.  However, the spins approximated by
vectors are not the original ones but are pseudo spins obtained
for quantum objects--spin singlet and triplet--and these plateaus
are not classical.

As has been mentioned above, the spin configurations obtained
in our approximate
method concern the pseudo spin defined by
triplet ($S^z=+1$) and singlet on a dimer bond.
In order to translate the classical pseudo-spin configuration into
the original $S=1/2$ one, it is convenient to consider the so-called
spin-coherent state which realizes almost `classical' states using
quantum states.
It is well-known that an $S=1/2$ coherent state
specified by a classical unit vector
\begin{equation}
\vec{\Omega} = (\cos \Phi \sin \Theta ,    \sin \Phi \sin \Theta,
\cos \Theta )
\end{equation}
is given by (aside from a phase factor coming from the
gauge degrees of freedom)
\begin{equation}
| \Phi, \Theta \rangle =
\be^{i \frac{1}{2}\Phi}\sin \left( \frac{\Theta}{2} \right)
|\mbox{singlet}\rangle +
\be^{-i\frac{1}{2}\Phi}\cos \left( \frac{\Theta}{2} \right)
|\mbox{triplet}\rangle \; .
\end{equation}
Plugging the expressions of $|\mbox{singlet}\rangle$ and
$|\mbox{triplet}\rangle$
in terms of original $S=1/2$, we obtain
\begin{eqnarray}
| \Phi, \Theta \rangle &=&
\be^{i\frac{1}{2}\Phi}\sin \left( \frac{\Theta}{2} \right)
\frac{1}{\sqrt{2}}(| \uparrow\downarrow\rangle
-| \downarrow\uparrow\rangle) \nonumber \\
& &+ \be^{-i\frac{1}{2}\Phi}\cos \left( \frac{\Theta}{2} \right)
| \uparrow\uparrow \rangle \; . \label{eqn:coherent}
\end{eqnarray}
Taking expectation values
\begin{eqnarray}
\langle \Phi, \Theta| S^{+}_{1} |\Phi, \Theta\rangle &= &
- \frac{1}{2\sqrt{2}} \be^{i \Phi} \sin \Theta \nonumber \\
\langle \Phi, \Theta| S^{+}_{2} |\Phi, \Theta\rangle &= &
 + \frac{1}{2\sqrt{2}} \be^{i \Phi} \sin \Theta \nonumber \\
\langle \Phi, \Theta| S^{z}_{1} |\Phi, \Theta\rangle &= &
\langle \Phi, \Theta| S^{z}_{2} |\Phi, \Theta\rangle =
\frac{1}{4} (1+\cos \Theta )  \; ,
\end{eqnarray}
we can see that in the presence of the superfluid LRO ($\Theta \ne 0$,
$\pi$)
the off-diagonal elements of
the original $S=1/2$ spins on a dimer bond align in an
antiparallel manner.

\begin{figure}[tbp]
  \begin{center}
    \leavevmode
    \epsfig{file=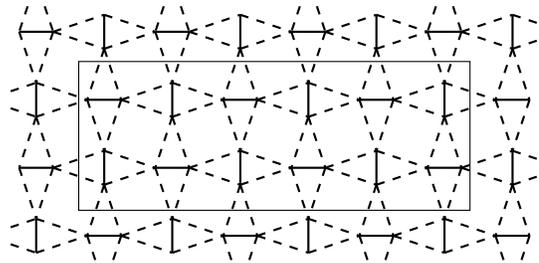,width=2.8in}
  \end{center}
  \caption{Finite-size ($N=24$) cluster that is used for the exact
   diagonalization study. }
  \label{fig:cluster}
\end{figure}

In order to check the accuracy of the strong coupling expansion and
the classical approximation,
we also studied a finite system of the original Shastry-Sutherland
model using exact diagonalization method.
We diagonalized 24-sites system with a periodic boundary
condition (Fig.\ \ref{fig:cluster}) that
matches with the 12-sublattice structure at magnetization
above $m/m_{\rm sat}=1/3$ (see section \ref{sec:supersolid}). The results
are shown in Fig.\ \ref{fig:mag_pro}.
Total behavior shows good agreement with the results from the strong
coupling expansion.

In the following, we discuss the nature of each phase and phase
diagram. 

%%%%%%%%%%%%%%%%%%%%%%%%%%%%%%%%%%%%%%%%%%%%%%%%%%%%
\subsection{SDW on plateau states}
\begin{figure}[tb]
  \begin{center}
    \leavevmode
    \epsfig{file=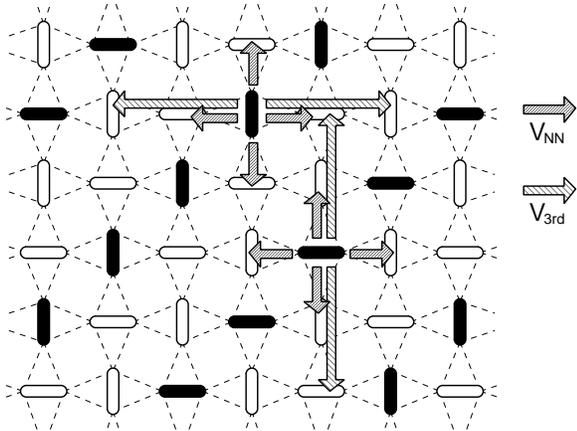,width=3.0in}
  \end{center}
  \caption{Stripe SDW structure at $m/m_{\rm sat}=1/3$, which is
    stabilized by the anisotropic repulsion $V_{\rm 3rd}$. Black bonds denote
    dimer triplet excitations and white ones dimer singlets.
    Arrows show range of repulsive interactions. }
  \label{fig:SDW0.3}
\end{figure}

%%%%%%%%%%%%%%%%%%%%%%%%%%%%%%%%%%%%%%%%%%%%%%%%%%%%
On the plateau states, dimer triplet excitations crystallize forming SDW
long-range orders, and there are spin gaps. 
The plateau states at $m/m_{\rm sat}=1/3$ and 1/2
have SDW with stripe and checkerboard structures, respectively, which
are shown in Figs.~6(a) and (b) in Ref.\ \onlinecite{MomoiT}. 
These configurations are
consistent with the insulating states naively expected from
the range and geometry of repulsive interactions.
The particles are perfectly closed packed at $m/m_{\rm sat}=1/2$ and 1/3
avoiding the repulsive interactions from 1st- and 2nd-order
perturbations, respectively. (See Fig.\ \ref{fig:SDW0.3} for the case
of $m/m_{\rm sat}=1/3$.)

%%%%%%%%%%%%%%%%%%%%%%%%%%%%%%%%%%%%%%%%%%%%%%%%%%%%%%%%%%%%
\subsection{Supersolid on non-plateau states}
\label{sec:supersolid}
%%%%%%%%%%%%%%%%%%%%%%%%%%%%%%%%%%%%%%%%%%%%%%%%%%%%%%%%%%%%
By applying a stronger magnetic field than the critical value,
the plateau states
continuously change to supersolid states in which superfluid
components appear and coexist with SDW of the plateau states.
Since the appearance of the superfluid component is accompanied by
the Goldstone bosons, the plateau gap collapses.
\begin{figure}[tb]
  \begin{center}
    \leavevmode
    \epsfig{file=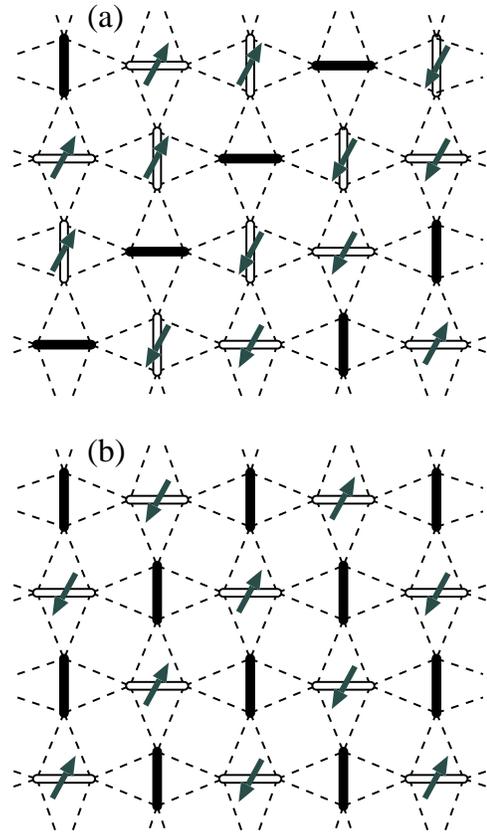,width=2.5in}
  \end{center}
  \caption{Spin configurations in supersolid phases slightly above
    $m/m_{\rm sat}=1/3$ (a) and 1/2 (b). Black bonds denote dimer
    triplet excitations and white ones with arrows denote linear
    combination of dimer triplet and singlet, which can be written as
    eq.\ (\ref{eqn:coherent}). The SDW components equal to
    those at
    $m/m_{\rm sat}=1/3$ and $1/2$, respectively. Superfluid components
    of magnetic excitations exist on white bonds, and arrows on white bonds
    denote phases $\Phi$ of particles (pseudo spins).
    The original spin configuration is recovered by eq.\ 
    (\ref{eqn:coherent}).}
  \label{fig:SS}
\end{figure}

\begin{figure}[tbp]
  \begin{center}
    \leavevmode
    \epsfig{file=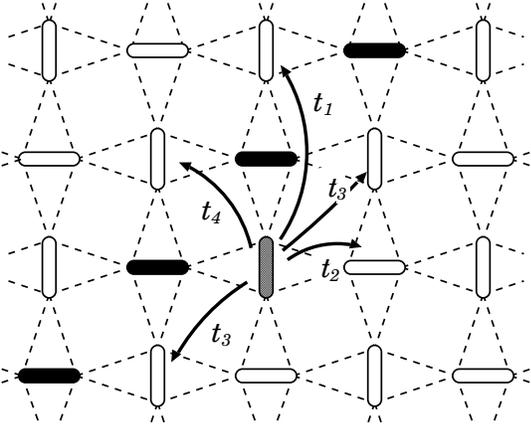,width=2.8in}
  \end{center}
  \caption{Hopping processes of a single dimer excitation (shaded one)
    near a line of triplet excitations (black bonds).
    $t_1 = (J^\prime)^2 n/4J + 3(J^\prime)^3 n/8J^2 $,
    $t_2 = -(J^\prime)^2 n/4J - 5(J^\prime)^3 n/16J^2 $,
    $t_3 = -(J^\prime)^3 n^2/8J^2 $, and
    $t_4 = (J^\prime)^3 n^2/8J^2 $. Here we
    assume the particle density on each black bond to be
    $n$. Matrices of hopping processes across the line ($t_1$, $t_4$)
    are all positive and other 
    ones are negative. }
  \label{fig:motion}
\end{figure}

Above the magnetization $m/m_{\rm sat}=1/3$,
particles with density 1/3 still form SDW with the stripe structure,
which equals to the one at $m/m_{\rm sat}=1/3$, and the rest of
particles become superfluid in `canals' between stripes.
(See Fig.\ \ref{fig:SS}(a).)
Particles can hop across a line of SDW with the help of correlated hopping
terms (see Fig.\ \ref{fig:motion}) and this hopping makes correlations
between superfluids in canals.
These extra particles can Bose-condense and the phases of the superfluid
particles align ferromagnetically inside an individual canal and
antiferromagnetically between canals.

The supersolid state above $m/m_{\rm sat}=1/2$ maintains the same SDW
as the $m/m_{\rm sat}=1/2$ case and have superfluid components as well.
Phases of superfluid components form a stripe
structure and the phase of one stripe aligns antiparallelly to those of
next ones. (See Fig.\ \ref{fig:SS}(b).)

In both supersolid phases,
a single dimer triplet can move by itself {\em assisted} by SDW LRO
and it behaves as an elementary particle of the superfluid.
The SDW forms a global network of crystallized particles and it helps
a dimer triplet to hop along the network owing to the
correlated hopping.
We roughly evaluate the hopping of an extra particle by treating
the stripe SDW as a classical background.
In Fig.\ \ref{fig:motion}, we show the matrix element of
single-dimer-triplet hopping along a stripe.
Note that triplet excitations are not confined in a one-dimensional
space (`canal') between two rows of particles.
A dimer triplet can hop both
parallel and perpendicular to the stripe.
The superfluid component of the system may behave as an
anisotropic two-dimensional Bose gas. The hopping
matrices inside of a stripe are negative, whereas those across a line
of triplet excitations are positive. This makes the phases of
superfluid component ferromagnetic inside the stripe and
antiferromagnetic between stripes.
A similar anisotropic superfluid was reported
for the Bose-Hubbard model with frustrating interactions
\cite{Otterlo-KHW-94}.

%%%%%%%%%%%%%%%%%%%%%%%%%%%%%%%%%%%%%%%%%%%%%%%
\subsection{Phase diagram}
%%%%%%%%%%%%%%%%%%%%%%%%%%%%%%%%%%%%%%%%%%%%%%%

\begin{figure}[tbp]
  \begin{center}
    \leavevmode
    \epsfig{file=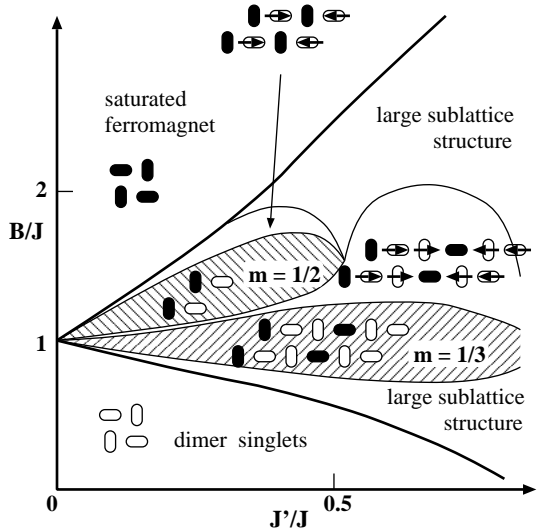,width=2.8in}
  \end{center}
  \caption{Phase diagram at $T=0$ with the parameter $J'/J$
    and the magnetic field $B/J$.
    The lower- and upper bold lines correspond to the 1st critical
    field and the saturation field, respectively.  For the latter,
    an analytic expression is available (see section V).
    Inserted figures denote density-wave order and phases of
    particles. Black bonds denote localized particles (dimer triplets)
    and white ones vacancies (dimer singlets). Arrows on dimer bonds
    denote phases of particles. }
  \label{fig:pd}
\end{figure}

We show phase diagram for $J'/J$ vs.\ $B/J$ at zero temperature in
Fig.\ \ref{fig:pd}, where the phase boundaries are not very accurate.
The plateau at $m/m_{\rm sat}=1/2$ appears
only in the region $0<J'/J<0.51$, and one at $m/m_{\rm sat}=1/3$ in
$0<J'/J<0.95$.
For large $J'/J$, insulating phases disappear and become superfluid
phases. This is because correlated hoppings are
dominant in the higher order terms and they become efficient for large
$J'/J$. The phase diagram may be not qualitatively accurate for large
$J'/J$, since our arguments are based on a strong coupling expansion.
Furthermore there is a possibility that the elementary particles change
to plaquette triplets\cite{KogaKawakami} for large $J'/J$, which is not
taken into account in our approximation.

Spin configurations at various values of magnetization are summarized
as follows:
\begin{itemize}
\item For $0< m/m_{\rm sat} < 1/3$, \\
spin configuration has large sublattice structures.
The system is highly frustrated and we sometimes reached to
different ground states with different magnetizations in the Monte
Calro method. We expect that weak interactions or quantum fluctuations
can drastically change the ground state properties. One
reason for this ambiguity may be that a single dimer triplet
is not an elementary particle at very low magnetization, but a bound
pair of triplets is. (See section \ref{sec:bound})
\item
At $m/m_{\rm sat}=1/3$,\\
there is only SDW LRO with stripe structure. (Fig.\ \ref{fig:SDW0.3})
\item
For $1/3< m/m_{\rm sat}< m_{c1}$ with $0.5<J'/J<0.8$, and for
$1/3< m/m_{\rm sat}< 1/2$ with $J'/J<0.5$, \\
the ground state shows supersolid with stripe structure. (Fig.\
\ref{fig:SS}(a)) 
\item
At $m/m_{\rm sat}=1/2$ for $J'/J<0.5$,\\
there is only SDW LRO with checkerboard structure.
(Fig.\ 6(b) in Ref.\ \onlinecite{MomoiT}.) The
results of the exact diagonalization (Figs.\ \ref{fig:mag_pro}(a) and (b))
indicate that this
phase seems to be more stable than in our approximation because of
quantum fluctuations. The phase boundary at $J'/J=0.51$ might move to a
larger value due to quantum effects.
One possible reason for this discrepancy is that 
our classical
approximation favors wave-like states (e.g.\ superfluid) against the
localized state.
\item
For $1/2 < m/m_{\rm sat}< m_{c2}$, if $J'/J<0.5$,\\
spins become supersolid with stripe structure. (Fig.\ \ref{fig:SS}(b))
\item
For $m_{c1}< m/m_{\rm sat}< 1$ with $0.5<J'/J$, and for
$m_{c2}<m/m_{\rm sat}<1$ with $J'/J<0.5$,\\
large sublattice structures, e.g. a helical structure, appear.
We will discuss nearly saturated region in Section \ref{sec:sat}.
A new quantum phase may appear in this region.
\end{itemize}

\section{Bound state of two dimer triplets near $m=0$}
\label{sec:bound}

In this section, we consider a striking effect of correlated
hopping on the dynamics of the Shastry-Sutherland model.
It is most clearly seen in the low-magnetization region where
the number of magnetically excited triplets is small.

First we suppose that there are only two excited triplets.
When they are far apart from each other, an individual
triplet can hardly hop\cite{MiyaharaUa}
and it gains little energy by moving on a lattice.
(This {\em almost localized} property was actually observed in
inelastic neutron-scattering experiments\cite{KageyamaB}.)
On the other hand, when the two are adjacent to each
other, the situation is completely different.
From the effective Hamiltonian, we can easily see
that {\em correlated}-hopping processes make
coherent motion of two triplets possible, where
a pair of triplet dimer excitations form a bound state
with $S^z=2$.
(In the same way we can easily derive various bound states with 
$S=0$, 1, 2 at zero magnetic field\cite{TotsukaMU}.
Here we only discuss the state with $S^z=2$, which becomes
dominant under the magnetic field.)
\begin{figure}[tbp]
  \begin{center}
    \leavevmode
    \epsfig{file=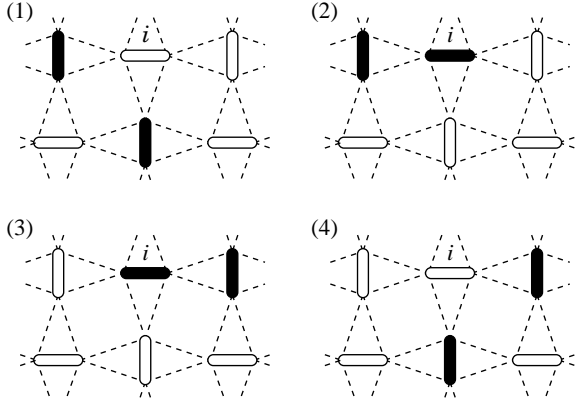,width=3.0in}
  \end{center}
  \caption{Four bases $|i, 1\rangle$ (1), $|i, 2\rangle$ (2),
    $|i, 3\rangle$ (3), and  $|i, 4\rangle$ (4) of a bound state of
    two dimer triplets. }
  \label{fig:base}
\end{figure}
\begin{figure}[tbp]
  \begin{center}
    \leavevmode
    \epsfig{file=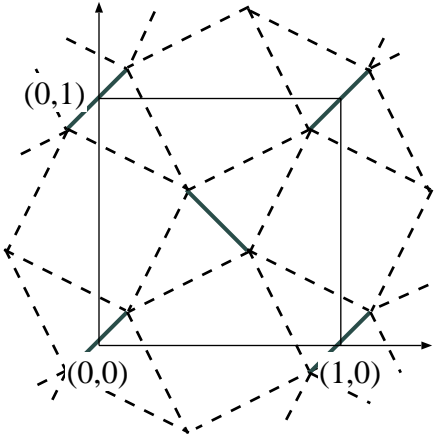,width=2.6in}
  \end{center}
  \caption{Chemical unit cell of the Shastry-Sutherland lattice. }
  \label{fig:unit_cell}
\end{figure}

Using the effective Hamiltonian, we can exactly show bound states.
Because of the correlating hoppings,
one triplet excitation exists necessarily in nearest
neighbor or next nearest neighbor sites of the other.
A little calculation shows that the hopping processes
are decomposed into the center-of-mass motion and the
relative motion, and that the latter is closed within the
four states shown in Fig.\ \ref{fig:base}.
Hence we can write the pair excited states as
\begin{eqnarray}
\sum_{i\in A} \exp(i {\bf P}\cdot {\bf r}_i) &\{& c_1({\bf P}) |i, 1\rangle
+c_2({\bf P}) |i, 2\rangle + c_3({\bf P}) |i, 3\rangle
\nonumber\\
& &+ c_4({\bf P}) |i, 4\rangle\},
\end{eqnarray}
where the two-dimensional momenta $\mbox{\bf P}=(p_{x},p_{y})$
are defined with respect
to the chemical unit cell of the Shastry-Sutherland lattice and unit
vectors are defined in Fig.\ \ref{fig:unit_cell}.
(In this sense, the readers should not be confused our
$\mbox{\bf P}$ with that used in refs.\
\onlinecite{Weihong-H-O-99} and \onlinecite{KageyamaB}.)
The energy spectra $\omega_{j}(\mbox{\bf P})$ ($j=1,2,3,4$)
of them are computed by diagonalizing the following hopping matrix:
\begin{equation}
\left(
\begin{array}{cccc}
2 V_0 + V_{\rm NNN} & J_{\rm NN} & J_{\rm NN}\be^{iP_y} & 0 \\
J_{\rm NN} & 2 V_0 + V_{\rm NN} & J_{3rd} & -J_{\rm NN}\be^{-iP_x} \\
J_{\rm NN}\be^{-iP_y} & J_{3rd} & 2 V_0 + V_{\rm NN} & -J_{\rm NN} \\
0      & -J_{\rm NN}\be^{iP_x} & -J_{\rm NN} & 2 V_0 + V_{\rm NNN}
\end{array}
\right),
\end{equation}
where
\begin{eqnarray}
V_0 &=& J-B-\dfrac{J'^2}{J}-\dfrac{J'^3}{2J^2}, \nonumber\\
J_{\rm NN} &=& \dfrac{J'^2}{4J}+\dfrac{5J'^3}{16J^2}, \nonumber\\
V_{\rm NN} &=& \dfrac{J'}{2}+\dfrac{J'^2}{2J}-\dfrac{J'^3}{8J^2}, \nonumber\\
V_{\rm NNN} &=& \dfrac{J'^3}{4J^2}, \nonumber\\
J_{\rm 3rd} &=& \dfrac{J'^2}{4J}+\dfrac{3J'^3}{8J^2} .
\end{eqnarray}
In addition, there is another type of correlated motion obtained
from the above one by reversing the space about $y$ axis.
Roughly speaking, this corresponds to the interchange of dimers A and B,
and its spectra are given by $\omega_{j}(-p_{x},p_{y})$.
These branches together with a dispersionless band of
a pair isolated triplets give the entire spectra of two-triplet sector.

\begin{figure}[tbp]
  \begin{center}
    \leavevmode
    \epsfig{file=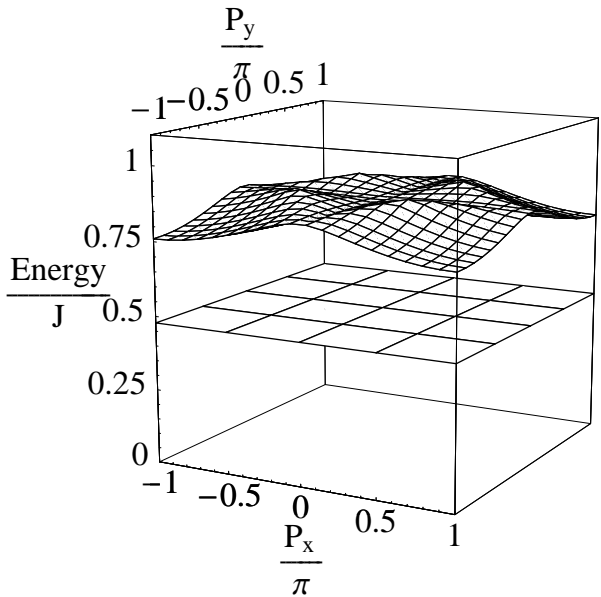,width=3.2in}
  \end{center}
  \caption{Energy spectrum of the effective Hamiltonian up to 3rd order
  at $J'/J=0.63$.   A two-triplet bound state consists of four
  branches and we show only the lowest one (see the upper band),
  which is apparently lower than the two-particle threshold.
  The spectrum of a single dimer triplet excitation ($V_0$) is
  also shown (lower band.  Note that it is dispersionless at this
  order.)}
  \label{fig:excitation}
\end{figure}
\begin{figure}[tbp]
  \begin{center}
    \leavevmode
    \epsfig{file=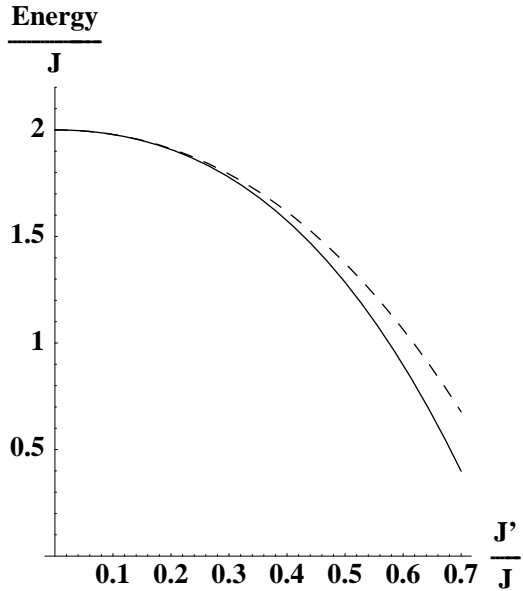,width=2.8in}
  \end{center}
  \caption{Energy of two dimer triplets for various $J'/J$.
    Dashed line denotes the energy of two non-interacting dimer
    triplets ($2V_0$) and solid line the
    energy of the bound state of two dimer triplets with the momentum
    ($\pi,\pi$). }
  \label{fig:energy}
\end{figure}

 Since these bound states can move because of correlated hopping
process, energy spectra of these states are dispersive. We show
only the lowest branch in Fig.\ \ref{fig:excitation}.
The lowest energy is given at $\mbox{\bf P}=(\pi,\pi)$ as
\begin{eqnarray}
\Delta E_2(B) &=& 2\,V_{0}  + \frac{1}{2}\left\{ -J_{\rm 3rd} + V_{\rm NN}
      + V_{\rm NNN} \right.
\nonumber\\
& &\left. - \sqrt{ \left(J_{\rm 3rd} - V_{\rm NN}
      + V_{\rm NNN}  \right)^{2} +16 J^{2}_{\rm NN} }  \right\}.
\label{b_energy}
\end{eqnarray}
On the other hand, two independent dimer triplet excitations
(scattering state) have
the dispersionless energy $2\Delta E_1(B)=2V_0$ (small dispersion appears
at 6th order and higher \cite{MiyaharaUa,Weihong-H-O-99}).
Expanding the right-hand side of eq.(\ref{b_energy}) in $J'/J$,
we can readily verify that there is a gain in kinetic energy of
$-J(J'/J)^3/4$ from the 2-particle threshold. Actually, the lowest
energy of the bound state is smaller than that of two
independent dimer triplets state for any $J'/J$ as shown in
Fig.\ \ref{fig:energy}.
For example, for $J'/J=0.68$ and $B=0$, the dispersion of
a bound state takes the minimal value $0.442J$
at $\mbox{\bf P}=(\pi,\pi)$,
whereas two independent dimer triplets have the energy $0.761J$ in
total. As is easily seen, two triplets combined to form
the $S=2$ bound states actually feel repulsive interactions between
each other (i.e. $V_{\rm NN}, V_{\rm NNN} >0$).
Since strong repulsion $V_{\rm NN}$ acts for a pair on
adjacent bonds, one may naively expect that such
bound motions are not energetically favorable.
However, we can take an optimal linear combination of
the four relative configurations $| i,\alpha\rangle$ with $\alpha=1,
2, 3, 4$, so that the
bound state may avoid the effect of $V_{\rm NN}$
(note that only $|i,2\rangle$ and $|i,3\rangle$ feel the repulsion
$V_{\rm NN}$) while gaining the kinetic energy by the coherent motion.
For large $V_{\rm NN}$ we can easily verify that
the effect of $V_{\rm NN}$ is canceled in $\Delta E_2$.
The energy gain due to the motion is larger than the cost
from repulsion, and hence relatively stable bound states are formed.

Next, we apply a magnetic field in the $z$-direction.
Because $S^{z}=2$ bound states have lower energy than two
unbound triplets, the bottom of
the bound states first touches the ground state when
the field is increased and one-triplet excitation still has a
finite energy gap at the critical magnetic field.
By increasing the magnetic field more than the critical field, a
macroscopic number of bound states condense
instead of the single dimer triplets.
Thus the non-plateau state at very low magnetization may be a
superfluid of bound states and is different from the almost localized
dimer triplet state discussed by Miyahara and Ueda\cite{MiyaharaUa}.
The critical magnetic field of magnetization process, where the
magnetization starts to increase from zero, corresponds
to half of the energy gap $\Delta E_2(0)$. If we regard the slow
increase of magnetization below $H_1=22.5$[T]\cite{Kageyama} as the
consequence of bound state condensation, we can estimate the gap
$\Delta E_2(0)$ as $51.0$[K] from magnetization
process\cite{Kageyama,Onizuka}.
On the other hand the one-triplet
energy gap $\Delta E_1(0)$ is estimated as $34.7$[K] from
susceptibility\cite{Kageyama}, 
inelastic neutron scattering\cite{KageyamaB}, and ESR\cite{Nojiri}.
If we set the
parameters as $J=81.4$[K] and $J'=53.5$[K], estimates of the
energy gaps, $\Delta E_1(0)$ and $\Delta E_2 (0)$, coincide
with the experimental results.

We expect that these bound states are destroyed at higher
magnetization. Indeed,
one-triplet excitations can move around above
$m/m_{\rm sat}=1/3$, as discussed in section~\ref{sec:supersolid}, and 
they can gain more kinetic energy than bound states.
There must be a transition of elementary particles from bound states
for very low magnetization to
one-triplet states for high magnetization. We can see this transition, 
if we neglect interactions
between particles and apply mean-field approximation to the correlated
hopping term. As particle density is
increased, one-particle hopping process effectively appears as
$(b_i^\dagger b_j+\mbox{\rm H.c.}) \langle n \rangle$ from the
correlated hopping term and then triplet particles gain kinetic energy.
In this rough estimation, a pair of unbound triplets have lower energy
than the bound state above $m/m_{\rm sat}=0.253$ for $J'/J=0.63$.
We can expect that bound states disintegrate above a finite value of
magnetization and one-triplet particles turn to elementary particles.
This estimate of critical magnetization can be highly modified by strong
correlation and readers should not consider the above value
seriously.

Finally we mention about crystallization of bound states.
One may expect bound states to crystallize at low magnetization,
but it will not occur.
If bound states tend to crystallize, bound states loose
kinetic energy and then they will be unbound by repulsive interaction
between triplets.

%%%%%%%%%%%%%%%%%%%%%%%
\section{Near Saturation}
\label{sec:sat}
%%%%%%%%%%%%%%%%%%%%%%%
There is another region where quantum effects manifest
themselves. In this section, we briefly discuss the region just below
the saturation field. In this region
the one-particle excitation can be obtained from the original
Shastry-Sutherland model without
any approximation.

The single-particle excitation over the fully polarized
ground state is given by a single flipped spin.
The dispersion of this excitation is readily computed by
diagonalizing the following hopping matrix
\begin{equation}
\left(
\begin{array}{cccc}
-\frac{1}{2}(J+4J^{\prime}) & \frac{J^{\prime}}{2}(1+ \be^{-ip_{x}})
& \frac{J}{2}\be^{-ip_{x}}\be^{-ip_{y}} &
\frac{J^{\prime}}{2}(1+\be^{-ip_{y}})  \\
\frac{J^{\prime}}{2}(1+\be^{ip_{x}}) &
-\frac{1}{2}(J+4J^{\prime}) &
\frac{J^{\prime}}{2}(1+\be^{-ip_{y}})  & \frac{J}{2} \\
\frac{J}{2}\be^{ip_{x}}\be^{ip_{y}}
 & \frac{J^{\prime}}{2}(1+\be^{ip_{y}})  &
-\frac{1}{2}(J+4J^{\prime})
& \frac{J^{\prime}}{2}(1+\be^{ip_{x}})   \\
\frac{J^{\prime}}{2}(1+\be^{ip_{y}})  & \frac{J}{2} &
\frac{J^{\prime}}{2}(1+\be^{-ip_{x}}) &
-\frac{1}{2}(J+4J^{\prime})
\end{array}
\right)   \; .
\end{equation}
In the above equation, momenta $(p_{x},p_{y})$ are defined
with respect to the chemical unit cell.
Reflecting the fact that a single unit cell contains four
spins, the spectrum consists of four bands.
Note that the four eigenvalues are invariant under
the point group $\mbox{D}_{\rm 2d}$:
\begin{equation}
(p_{x},p_{y}) \mapsto (p_{y},p_{x})
\quad \mbox{and} \quad
(p_{x},p_{y}) \mapsto (-p_{x},p_{y})  \; .
\end{equation}
In the dimer limit $J^{\prime}/J \ll 1$, the lower two
correspond to energy of a singlet particle on dimer bonds, where the
number two comes from the two mutually orthogonal dimer bonds,
and the higher two to a triplet ($S^{z}=0$) one; our approximation
in section \ref{sec:effec_H} corresponds to neglecting the latter
as higher-lying.

Although the expression of the dispersion is rather complicated,
we can locate the position of the band minima in the momentum
space, which is relevant in determining the structure in the vicinity
of saturation.
For $0 \leq J^{\prime}/J <1$, the lowest band takes a minimal value
$-J(1+J^{\prime}/J)^{2}$ {\em on a closed curve}
\begin{equation}
\cos p_{x} + \cos p_{y} =
2\left( \frac{J^{\prime}}{J}\right)^{2}  \; .
\end{equation}
This implies that magnetization saturates at
\begin{equation}
H_{\rm c}= J \left( 1+\frac{J^{\prime}}{J} \right)^{2} \; .
\end{equation}

Because of a dispersionless mode on the closed curve,
the density of states (DOS) near
saturation magnetization is like 1D and different from
the usual 2D one. Hence the singularity of magnetization curve near
saturation magnetization would be 1D-like,
i.e., $|m_{\rm c}-m|\sim \sqrt{|H_{\rm c} - H|}$.
This behavior can be seen in the results of the exact
diagonalization in Fig.\ \ref{fig:mag_pro}.

The location of the dispersion minimum
$\mbox{\boldmath$p$}_{\rm min}$ together with the corresponding
eigenvector determines the spin structure at the semiclassical level.
Usually spin states just below the critical field are correctly given
by the classical model \cite{Batyev-B-84,Gluzman}.
Detailed analyses of the wave functions reveal that
the fourfold-degenerate classical helical order, which was
pointed out by Shastry and Sutherland\cite{ShastryS},
corresponds to the four apexes $(\pm p_{\rm max},0),
(0,\pm p_{\rm max})$ ($p_{\rm max}=2\cos^{-1}(J^{\prime}/J)$)
of the closed curve.
The fate of the classical helical LRO when the quantum effects are
fully taken into account is unclear at present.
The four ground states with helical LRO are connected by a gapless
line, which means stiffness of helical order is vanishing.
We hence expect that quantum fluctuations
destroy the classical LRO.

%One may expect that bound states of singlet-pairs realize 
%near saturation magnetization, but this would not be stable. Applying the 
%particle-hole transformation to the effective Hamiltonian 
%(\ref{eqn:eff_Ham}), we can
%derive effective Hamiltonian for the singles. This Hamiltonian
%contains hopping terms like $(b_i^\dagger b_j + h.c.)(1-n_k)$,
%where $b^\dagger$ and $n$ are the creation and number
%operators of singlets. This term makes
%a particle(singlet) hop, but if another particle exists on the site 
%$k$ this hopping process does not occur. To avoid loss of kinetic
%energy, particles prefer separated each other. Furthermore, diagonal 
%interaction
%terms contain repulsions at lowest order of $J^\prime/J$.
%Though attractive interactions appear at second order, this term is
%weaker than hopping and repulsive terms and hence it will not stabilize
%the bound state.

On the other hand, for $J^{\prime}/J \geq 1$
the saturation field $H_{\rm c}=4J^{\prime}$ becomes
$J$-independent and
the band minimum shrinks to a single point
$\mbox{\boldmath$p$}=(0,0)$;
the spin structure realized in the vicinity of saturation
is the classical staggered one where spins connected by
$J^{\prime}$-bonds align antiferromagnetically in the
$xy$-plane.  Correspondingly, the transition to saturation
is the same as that of 2D superfluid\cite{DFisherH}.

\section{Critical Phenomena}
\label{sec:critical}
According to an analogy to many-particle theories, a plateau 
state corresponds to a SDW insulating state and gapless ones to
supersolids.
As the plateau states at $m/m_{\rm sat}=0$, $1/3$ and $1/2$ collapse by
increasing the magnetic field, 
superfluid components appear, whereas the SDW structure still remains
in each phase.
Contrary to one dimension, superfluid LRO is accompanied by the
gapless Goldstone mode in our two-dimensional case.
Hence, collapse of plateau occurs at the same point with the
onset of superfluidity component. On the other hand if we look at the
lower phase boundaries of the SDW
phases, SDW structures change at the transition points in our
approximation.
In this case the phase transitions are presumably of first order. 

Now let us consider the case of 2nd order transition. 
We can imagine two different situations.
The first one is (i) the transitions driven by changing
the external field while the coupling $J^{\prime}/J$ is kept
fixed. This transition is seen in the actual magnetization process.
This type of field-induced transitions has a close relationship
to the filling-control insulator-to-superfluid transitions in Bose
systems, and the methods used there can be
imported to our case.
The second one is (ii) continuous quantum phase transitions
occurring with fixed values of magnetization.
As the limit $J^{\prime}/J \nearrow \infty$ is
approached, the Shastry-Sutherland model reduces to
the ordinary $S=1/2$ square-lattice Heisenberg antiferromagnet,
where no magnetization plateau appears. Hence magnetization plateaus
vanish at some critical values of $J^{\prime}/J$ and are superseded by
supersolid (or, superfluid) phases.
%In the above case, plateau-superfluid transitions are brought about
%by controlling quantum fluctuation
%between two competing tendencies--crystallization (plateaus)
%and delocalization (superfluids).
In the present model, there is no particle-hole
symmetry around insulating phases apparently. We hence conclude that
the above two transitions have the same universality class.

%%%%%%%
%\subsection{Coupling-control transitions}
%%%%%%%%
%We begin with the case (i).
First of all, we have to keep in mind that because of the
special geometry of the Shastry-Sutherland model there is no {\em a priori}
reason for believing that the system is described by the
ordinary Bose liquid with a well-defined one-particle
dispersion $\varepsilon(p) \sim p^{2}$.
Actually, the low-order effective Hamiltonian
(\ref{eqn:eff_Ham}) lacks the one-particle part.
However, we have seen in section IV that this leads
to the formation of dispersive two-triplet bound states
in the low-field region, and in section \ref{sec:supersolid} 
that SDW structure makes
one particle dispersive in the supersolid phase around 
$m/m_{\rm sat}=1/3$ and 1/2. 

Near the phase boundaries, the superfluid amplitude is small
and we can map the problem onto the effective Ginzburg-Landau model
described by the superfluid order parameter \cite{Doniach-81}.
%We assume that low-energy property around the critical point is
%effectively described with two-dimensional interacting boson system at
%low-density region.
(These Bose particles are not necessarily dimer triplet excitations, 
but they can be plaquette triplet states of two dimer triplets
or flipped spin states\cite{MomoiT}.)
This enables us to conclude that superfluid-onset transition
would be described by the (2+1)-dimensional classical
XY model when the particle-hole
symmetry exists and by the $z=2$ mean-field-like transitions
when it does not\cite{MFisherWGF,DFisherH}.
In our case, the effective boson model in section II does not
have particle-hole symmetry and we hence conclude
that the plateau transition is of the dynamical exponent
$z=2$ and behaves as
\begin{eqnarray}
|m-m_{\rm c}| &\sim& |H-H_{\rm c}|\log (C/|H-H_{\rm c}|)
\nonumber\\
& &\mbox{\hspace{3cm}\rm (2D system).}
\end{eqnarray}
in two dimensions. Here $m_{\rm c}$ and $H_{\rm c}$ denote the critical
magnetization and field at the plateau transition. In the real
material, there are weak interactions between two-dimensional layers and
these interactions will push the system above the upper critical
dimension, i.e.
\begin{eqnarray}
|m-m_{\rm c}| &\sim& |H-H_{\rm c}|
\nonumber\\
& &\mbox{\hspace{1cm}\rm (3D and quasi-2D systems).}
\label{critical3d}
\end{eqnarray}
Note that these forms are quite different from that in one-dimensional 
systems,\cite{Totsuka}
$|m-m_{\rm c}| \sim \sqrt{|H-H_{\rm c}|}$.

%(This should be contrasted with the coupling-control plateau-superfluid
%transition occurring for the half plateau of the Heisenberg model
%on a 1/5-depleted square lattice \cite{MomoiT}.)

%The dynamical exponent $z$ takes the value unity and
%the correlation length diverges like $\xi_{\rm space} \sim
%\xi_{\rm time} \sim |(J^{\prime}/J)-(J^{\prime}/J)_{\rm c}|^{\nu}$
%with the exponent \cite{LeGuillou-Z-80} $\nu\approx 0.669$.

%%%%%%%
%\subsection{Field-induced transitions}
%%%%%%%%
%Next, we proceed to the field-induced transitions.

Here we also give a comment about the case the kinetic term of Bose
particles does not behave as $\varepsilon(p) \sim p^{2}$. 
According to the scaling argument of Ref.\ \onlinecite{MFisherWGF},
exponents of this kind of transitions should satisfy the relation
\begin{equation}
z \nu =1 \; .
\end{equation}
If the one-particle dispersion is of the form
$\varepsilon(p) \sim p^{l}$ ($l$ denotes an even integer),
the standard power-counting argument shows that
the upper critical dimensions are given by $d_{\rm u}=l$.
For $d\le l$, we obtain $|m-m_{\rm c}| \sim |H-H_{\rm c}|^{d/l}$.

For the case that SDW structure changes at phase boundary,
e.g. the boundary between
the plateau state at $m/m_{\rm sat}=1/2$ and the supersolid one
below $m/m_{\rm sat}=1/2$, the system shows a 1st order phase
transition in our approximation.
We don't find any incommensurate phase between them.
To take into account the possibility of any 2nd-order phase transition
or incommensurate phase, we need to consider the effect of quantum
fluctuations more seriously.

\section{Discussions and future problems}

To summarize, we studied the magnetic behavior of the
Shastry-Sutherland model using strong-coupling expansion. 
Magnetic excitations show insulator-supersolid transitions at
magnetization $m/m_{\rm sat}=1/3$ and 1/2, and thereby create
magnetization plateaus. 
Magnetization curve
obtained near $m/m_{\rm sat}=1/3$ looks similar to the experimental 
result.\cite{Onizuka} 

At zero magnetization, 
bound states of triplet excitations are formed by the correlated
hopping process. Above the 
critical field, quintuplet ($S=2$) bound states become elementary
particles in the ground state and they condense. Whereas, for 
large magnetization, the bound states are destroyed in the ground
states and triplet excitations become elementary particles. Triplet
excitations are essential for the plateau transitions at $m/m_{\rm
sat}=1/3$ and 1/2.
In the experiments, it is unclear in which region bound states appear
as elementary particles.
One possibility is that they appear in the tail of the
magnetization process below $m/m_{\rm sat} \approx 0.025$.
In this region, slope of the magnetization curve is different from
that of rest parts.\cite{Kageyama,Onizuka}
Further detail analysis are needed on the nature of quasiparticles.

In the present analysis, we did not find the plateaus at
$m/m_{\rm sat}=1/8$ and 1/4.
The mechanism of stabilizing these plateaus is
not yet clear. It may be natural to believe that dimer triplet excitations 
are crystallized by longer-range repulsive interaction. ($S=2$ bound states
cannot crystallize because of a special origin of the binding energy
as we discussed in section \ref{sec:bound}.)
We can consider two origins for the repulsions as follows:
\begin{itemize}
\item[1)] Though we cut the perturbation series at 3rd order, 
the higher-order expansions can produce longer-range repulsions
between particles. These repulsive interactions
may induce crystallization at low magnetization.
\item[2)] Longer-range repulsions can come from other
antiferromagnetic spin interactions in the original spin model, which
have not been accounted for in the pure Shastry-Sutherland model.
If we treat the
antiferromagnetic interactions between next-neighbor dimer bonds, 
we can produce crystallization at low magnetization.
For example, M\"{u}ller-Hartmann et al.\ demonstrated 
the appearance of 1/4-plateau considering another spin interaction, 
which acts between nearest-neighbor dimers.\cite{MuellerH}
\end{itemize}
It is unclear which spin interaction is important in the
real material.
We need first-principle calculation to estimate exchange couplings.
We also need to keep in mind that the real material is close to the
plaquette singlet phase. Under the magnetic field, if this phase
becomes more stable than the dimer singlet state and plaquette
triplets become elementary particles, plaquette triplets may
crystallize and hence create magnetization plateaus at low
magnetization as discussed in refs.\ \onlinecite{MomoiT,KogaKawakami}.
This should be considered in a future.

\acknowledgements

We would like to thank the late Dr.\ Nobuyuki Katoh for stimulating
discussions at the beginning of this study. We also thank Hiroshi
Kageyama, Norio Kawakami, Kenn Kubo, Hiroyuki Nojiri, and Kazuo Ueda
for useful comments and 
discussions. We also acknowledge Hiroshi Kageyama and Hiroyuki Nojiri
for showing us their experimental results before publication. 
One of us (TM) acknowledges the condensed matter
theory group in the Harvard University for kind hospitality. 
KT is financially supported by Inoue fellowship. 

%%%%%%%%%%%%%%%%%%%%%%%%%%%%%%%%%%%
\appendix
%%%%%%%
\section{Effective Hamiltonian}
%%%%%%%%%%%%%%%%%%%%%%%
In this appendix, we briefly explain how to obtain
the effective Hamiltonian in the framework of
degenerate perturbation.
We suppose that the ground states of the unperturbed
Hamiltonian (${\cal H}_{0}$) are degenerated and
we diagonalize these
degenerate ground states by degenerate perturbation.
Let the operators $P_{0}$ and $Q_{0}$
be the projection operators
onto the degenerate (unperturbed) ground-state sector and
its complement, respectively.   In addition, we define
a projection $P$ onto the perturbed ground-state sector.

According to ref.\ \onlinecite{Kato-76},
the problem of degenerate
perturbation reduces to solving the following problem
\begin{equation}
P_{0}{\cal H}P P_{0} | \mbox{G.S.; }\alpha \rangle =
E_{\alpha}K | \mbox{G.S.; }\alpha \rangle \; ,
\label{eqn:Rellich-Kato}
\end{equation}
where the hermitian operator $K$ is defined by
\begin{equation}
K \equiv P_{0}P P_{0} \; .
\end{equation}

However, this form is not so convenient to our purpose
because it does not take the form of the ordinary eigenvalue
problem.   A trick invented by Bloch\cite{Bloch-58} solves this difficulty.
The key is to introduce a new operator $U$ through the
following relation
\begin{equation}
P P_{0} = U K \; .
\end{equation}
The operator $U$ can be expanded as
\begin{equation}
U = \sum_{n=0}^{\infty} \lambda^{n} U^{(n)} \; ,
\end{equation}
where the $n$-th order coefficients $U^{(n)}$ are given
by
\begin{eqnarray}
U^{(0)} &=& P_{0} \nonumber \\
U^{(1)} &=& S V P_{0} \nonumber \\
U^{(2)} &=& \left(
S VS V+S^{2} V S^{0} V \right)P_{0} \nonumber  \\
U^{(3)} &=& \left(   SVSVSV + S^{3}VS^{0}VS^{0}V + S^{2}VSVS^{0}V \right.
\nonumber\\
& &\left. + SVS^{2}VS^{0}V + S^{2}VS^{0}VSV \right)P_{0}
\nonumber \\
& & \cdots
\end{eqnarray}
In the above equation, we have used a short-hand notation
$S^{k}$ defined by
\begin{equation}
S^{k} \equiv
\left\{
\begin{array}{ll}
-P_{0} & \mbox{ for } k=0 \\
\frac{1}{(E-{\cal H}_{0})^{k}}Q_{0} & \mbox{ for } k\geq 1 \; .
\end{array} \right.
\end{equation}

Then, eq.\ (\ref{eqn:Rellich-Kato}) can be recasted as
\begin{equation}
\left( P_{0}{\cal H}U - E_{\alpha} \right) K
|\mbox{G.S.; }\alpha \rangle =0 \; .
\end{equation}
This is a usual eigenvalue problem and
the matrix we have to diagonalize is finally given by
\begin{eqnarray}
{\cal H}_{\rm eff} &=& P_{0}{\cal H} U \nonumber \\
&=& E_{0} + \lambda P_{0}VP_{0}
+ \lambda^{2}P_{0}V\frac{1}{E_{0}-{\cal H}_{0}}Q_{0}VP_{0}
\nonumber \\
& & \; \;  + \lambda^{3} \left[
P_{0}V\frac{1}{E_{0}-{\cal H}_{0}}Q_{0}V
\frac{1}{E_{0}-{\cal H}_{0}}Q_{0}V P_{0} \right.
\nonumber\\
& &\left. -P_{0}V\frac{1}{(E_{0}-{\cal H}_{0})^{2}}Q_{0}VP_{0}VP_{0}
\right] + \cdots .
\end{eqnarray}
Note that non-hermitian terms appear in general
when we proceed to terms higher than second order.
Reality of the eigenvalues is no longer guaranteed.
To remedy this shortcomings, we have used the average of
${\cal H}_{\rm eff}$
and ${\cal H}^{\dagger}_{\rm eff}$, which is now hermitian.
%%%%%%%%%%%%%%%%%%%%%%%%%%%%%%%%%

\end{document}